\documentclass[aps,prd,reprint,groupedaddress,nofootinbib]{revtex4-1}

\usepackage{graphicx}
\usepackage{longtable}

\usepackage{relsize}
\def\Babar{{\mbox{\slshape B\kern-0.1em{\smaller A}\kern-0.1em B\kern-0.1em{\smaller A\kern-0.2em R}}}}

 \usepackage{amssymb} \usepackage{amsmath}
\usepackage{color}
\usepackage{url}
\usepackage[bookmarks, breaklinks, colorlinks,urlcolor=black, citecolor=red, 
linkcolor=blue]{hyperref}

\def\bar {\overline}

\def\bea {\begin{eqnarray}}
\def\eea {\end{eqnarray}}

\def\beq{\begin{equation}}
\def\eeq{\end{equation}}
\def\barr{\begin{array}}
\def\earr{\end{array}}

\def\ltap{\raisebox{-.4ex}{\rlap{$\sim$}} \raisebox{.4ex}{$<$}} 
\def\gtap{\raisebox{-.4ex}{\rlap{$\sim$}} \raisebox{.4ex}{$>$}}

\def\gev{\ensuremath{\,\mathrm{Ge\kern -0.1em V}}}
\def\tev{\ensuremath{\,\mathrm{Te\kern -0.1em V}}}
\def\fb{\ensuremath{\,\mathrm{fb}}}

\usepackage{cancel}
 \usepackage[normalem]{ulem}
\usepackage{color}

\def\lapp{\mathrel{\rlap{\raise.5ex\hbox{$<$}}
                    {\lower.5ex\hbox{$\sim$}}}}
\def\gapp{\mathrel{\rlap{\raise.5ex\hbox{$>$}}
                    {\lower.5ex\hbox{$\sim$}}}}
\preprint{IP/BBSR/2017-17}

\begin{document}

\title{Non-Standard Neutrino Interactions : Obviating Oscillation Experiments}

\author{Debajyoti Choudhury$^1$, Kirtiman Ghosh$^2$ and 
  Saurabh Niyogi$^3$}
\affiliation{$^1$Department of Physics and Astrophysics, University of Delhi, 
Delhi 110007, India\\
$^2$Institute of Physics, Bhubaneswar 751005 \& HBNI, Mumbai, India\\
$^3$\hbox{Department of Physics, Gokhale Memorial Girls' College, Harish Mukherjee Road, 
Kolkata 700020, India}
}
\email{$^1$debajyoti.choudhury@gmail.com\\
$^2$kirti.gh@gmail.com\\
$^3$saurabhphys@gmail.com}


\begin{abstract}
 {Searching for non-standard neutrino interactions, as a means for
  discovering physics beyond the Standard Model, has one of the key
  goals of dedicated neutrino experiments, current and future. We
  demonstrate here that much of the parameter space accessible to such
  experiments   {is} already ruled out by the 
  RUN II data of the Large Hadron Collider experiment.}
\end{abstract}
\maketitle

Precision measurements of the neutrino mixing parameters, made over
the past few decades has significantly shortened the list of
unanswered questions in the standard scenario to just the issues of
the neutrino mass hierarchy i.e., ${\rm sign}(\delta m^2_{31})$, the
CP phase and the correct octant for the mixing angle $\theta_{23}$.
While the simplest way to generate neutrino masses is to add right
handed neutrino fields to the Standard Model (SM) particle content, it
is hard to explain  their extreme smallness. Several scenarios
going beyond the SM have been proposed to this end, often tying up
with other unanswered questions such as (electroweak)
leptogenesis~\cite{Pilaftsis:2005rv, Choudhury:2012zz}, neutrino
magnetic
moments~\cite{Barbieri:1988fh,Babu:1989tq,Choudhury:1989pw,Choudhury:1991sd},
neutrino condensate as dark
energy~\cite{Bhatt:2009wb,Wetterich:2014gaa}. An agnostic alternative
is to add dimension-five terms consistent with the symmetries and
particle content of the SM, which naturally leads to desired tiny
Majorana masses for the left-handed neutrinos.   {Irrespective of
  the approach,} once new physics is invoked to explain the non-zero
neutrino masses, it is unnatural to exclude the possibility of
non-standard interactions (NSI) as well. Indeed, NSI has been studied
in the context of atmospheric
neutrinos~\cite{Fornengo:2001pm,Huber:2001zw,GonzalezGarcia:2004wg,GonzalezGarcia:2011my,Esmaili:2014ota,Chatterjee:2014gxa},
CPT violation~\cite{Chatterjee:2014oda, Datta2004356}, violation of
the equivalence principle~\cite{Esmaili:2014ota}, large extra
dimension models~\cite{Esmaili:2014esa} {,} sterile
neutrinos~\cite{Esmaili:2012nz,Esmaili:2013cja,Esmaili:2013vza} and
collider experiments
\cite{Davidson:2011kr,Friedland:2011za,Davidson:2011xz,Franzosi:2015wha}.

At sufficiently low energies, a wide class of new physics scenarios
can be parameterised, in a model independent way, through the use of
effective four-fermion interaction terms. While these, in general,
would incorporate both charged-current (CC) and neutral-current (NC)
interactions, we shall confine ourselves largely to the latter (coming
back to the former only later). The dimension-6 
neutrino-quark interactions can, then, be expressed, in terms of the 
chirality projection operators $P_X \, (X = L,R)$, as
\begin{equation}
{\cal L}_{4}=- 2\sqrt 2 \,G_F \, \epsilon_{\alpha \beta}^{qX} \,
   \left(\bar q \gamma_\mu P_X q\right)
   \left(\bar \nu_\alpha \gamma^\mu P_L \nu_\beta\right)+ H.c.,
\label{4fermi}
\end{equation}
where $\alpha,\beta$ denote the neutrino flavours, $q$ is a quark
field, and $\epsilon_{\alpha \beta}^{qX}$ are arbitrary constants,
presumably $\ltap {\cal O}(10^{-2})$. It should be noted that
flavour-changing currents are allowed at the neutrino-end but not for
the quarks, with this restriction being imposed to evade the strong
bounds from decays such as $K \to \pi \nu\bar\nu$ or $B \to K
\nu\bar\nu$.  While this might seem an unnatural choice (note also
that analogous currents involving the charged leptons would,
typically, be subjected to even stronger constraints), the inclusion of
such  {flavour-changing} neutrino currents is not crucial to the
main import of this paper.

Neutrino oscillation experiments can probe such NSI by exploiting the
interference with the SM amplitude, with the NC interactions altering
the refractive index, as evinced by the far detectors. The excellent
agreement of data with the standard flavour conversion paradigm
implies that reasonably strong constraints are already in place with
these slated to improve considerably in the next-generation
experiments.

At the LHC, operators as in eq.(\ref{4fermi}) would lead to a change
in the rates for final states comprising a hard jet and missing
energy.  For $q = u, d$, this would be dominated by parton-level
processes such as $q + g \to q + \nu + \bar\nu$ and $q + \bar q \to g
+ \nu + \bar\nu$.  With the (anti-)neutrinos going undetected,
  {different choices of $\alpha,
  \beta$} would lead to essentially the same observables, and are,
hence, indistinguishable from each other. While the aforementioned
subprocesses dominate at the partonic level, the detector could also
register multiple jets (with missing energy) accruing from initial and
final state radiations, hadronization etc. Indeed, such processes have
been studied extensively \cite{ATLAS-CONF-2017-060,Aaboud:2017phn,Sirunyan:2017hci,Sirunyan:2017jix} 
as a search tool for new physics scenarios
such as supersymmetry, extra dimensions as well as generic Dark Matter
models.

To generate events at the LHC, we have incorporated the 4-fermi
operators ({of} eq.(\ref{4fermi})) in FeynRules
(v2.3.13) \cite{Alloul:2013bka,Christensen:2008py} to generate model
files for MadGraph5\_aMC@NLO (v2.2.1)~\cite{Alwall:2014hca}. In order
to compute the cross sections, we have used  {the} NNPDF23lo1
parton distributions~\cite{Ball:2012cx} with the factorization and
renormalization scales kept fixed at the central $m_T^2$ scale after
$k_T$-clustering of the event.  Initial and final state radiation,
showering and hadronization were simulated with PYTHIA
6.4~\cite{Sjostrand:2006za}.  The reconstruction of physics objects
(jets, leptons, $\cancel E_T$ etc.)  was done in accordance with the
prescription of the ATLAS monojet + $\cancel E_T$
analysis~\cite{Aaboud:2017phn}. We have used FastJet
\cite{Cacciari:2011ma} and the anti--$k_T$ jet clustering
algorithm~\cite{Cacciari:2008gp} with a radius parameter of $0.4$ for jet reconstruction . While
only jets with $p_T>20~\gev$ and $|\eta|<2.8$ are retained, electron
(muon) candidates are required to have $p_T > 20~(10)~\gev$ and $|\eta|
< 2.47~(2.5)$.  The discarding any putative jet lying within a distance
$\Delta R = \sqrt{\Delta \eta^2 + \Delta \phi^2} < 0.2~{(0.4)}$ of an
electron {(muon)} candidate resolves overlaps. {Moreover, for events with} $0.2<\Delta
R_{ej}<0.4$,  {the electron is removed} as it is likely
 {to have emanated} from a semileptonic
b-hadron decay.  The missing transverse momentum is reconstructed
using all energy deposits in the calorimeter (including unassociated
calorimeter clusters) up to pseudorapidity $|\eta| < 4.9$.  Only
events with zero leptons, $\cancel E_T > 250~\gev$ and atleast one jet
(satisfying the aforementioned preselection criteria) are selected for
further analysis.

A monojet-like final state topology demands a leading jet with $p_T >
250~\gev$ and $|\eta| < 2.4$. On the other hand, a maximum of four jets
with $p_T > 30~\gev$ and $|\eta| < 2.8$ are allowed. Additionally, to
reduce the multijet background contribution where a large $\cancel
E_T$ can originate from jet energy mismeasurement, each of the jets
must satisfy a azimuthal separation criterion of $\Delta
\phi(jet,\vec{\cancel E_T}) > {0.4}$. Subsequently, different signal regions
(IM1--SR10) are defined, in accordance with the  
ATLAS monojet-like selection criteria~\cite{Aaboud:2017phn}, 
 with progressively increasing thresholds for
$\cancel E_T$. These are summarized in Table~\ref{monojet_selection}. 
\begin{table}[h]
\begin{center}
\begin{tabular}{||c||c|c||c|c||}
\hline \hline
SR & $\cancel E_T$ (GeV) & $\sigma_{\rm{obs}}^{95}$ (fb) &   \multicolumn{2} {|c||} {$\sigma_{\rm{exp}}^{95}$ (fb)}
\\
\hline
   &                     &     36.1 fb$^{-1}$          & 100 fb$^{-1}$ & 300 fb$^{-1}$ \\
\hline\hline
IM1 & $ > 250$  & 531 & $160^{+80}_{-43}$ & $80^{+41}_{-31}$ \\ 
\hline
IM2 & $ > 300$  & 330 &  $94^{+48}_{-37}$ & $47^{+24}_{-18}$ \\
\hline
IM3 & $ > 350$  & 188 &  $52^{+26}_{-21}$ & $26^{+13}_{-10}$ \\
\hline
IM4 & $ > 400$  & 93  &  $28^{+13}_{-11}$ & $14^{+7}_{-5}$ \\
\hline
IM5 & $ > 500$  & 43  &  $10^{+5}_{-4}$ & $5.1^{+2.4}_{-1.9}$\\
\hline
IM6 & $ > 600$  & 19  &   $4.8^{+2.2}_{-1.8}$ & $2.5^{+1.1}_{-0.9}$\\
\hline
IM7 & $ > 700$  & 7.7 &  $2.9^{+1.3}_{-1.0}$ & $1.5^{+0.6}_{-0.5}$\\
\hline
IM8 & $ > 800$  & 4.9 &  $1.8^{+0.8}_{-0.6}$ & $0.9^{+0.4}_{-0.3}$\\
\hline
IM9 & $ > 900$  & 2.2 &  $1.2^{+0.5}_{-0.4}$ & $0.6^{+0.3}_{-0.2}$\\
\hline
IM10 &$ > 1000$ & 1.6 & $0.8^{+0.3}_{-0.3}$ & $0.4^{+0.2}_{-0.1}$\\
\hline
\end{tabular}
\end{center}

\caption{The different monojet-like signal regions as defined by the
  ATLAS collaboration~\cite{ATLAS-CONF-2017-060,Aaboud:2017phn},  {and
    their} corresponding 95\% CL upper limits  {(for $36.1\fb^{-1}$ data)}
  on the 
  cross section ($\sigma_{\rm obs}^{95}$)  {due to all BSM
    effects}.  {Assuming that the
    agreement between the data and SM would persist,
    $\sigma_{\rm{exp}}^{95}$ represent the expected 95\% CL upper
    limits for integrated luminosities of 100 and $300\fb^{-1}$.}}
\label{monojet_selection}
\end{table}

\begin{figure}[h]
\begin{center}
\vskip -15pt
\includegraphics[angle =0, width=0.495\textwidth, height=0.3\textwidth]{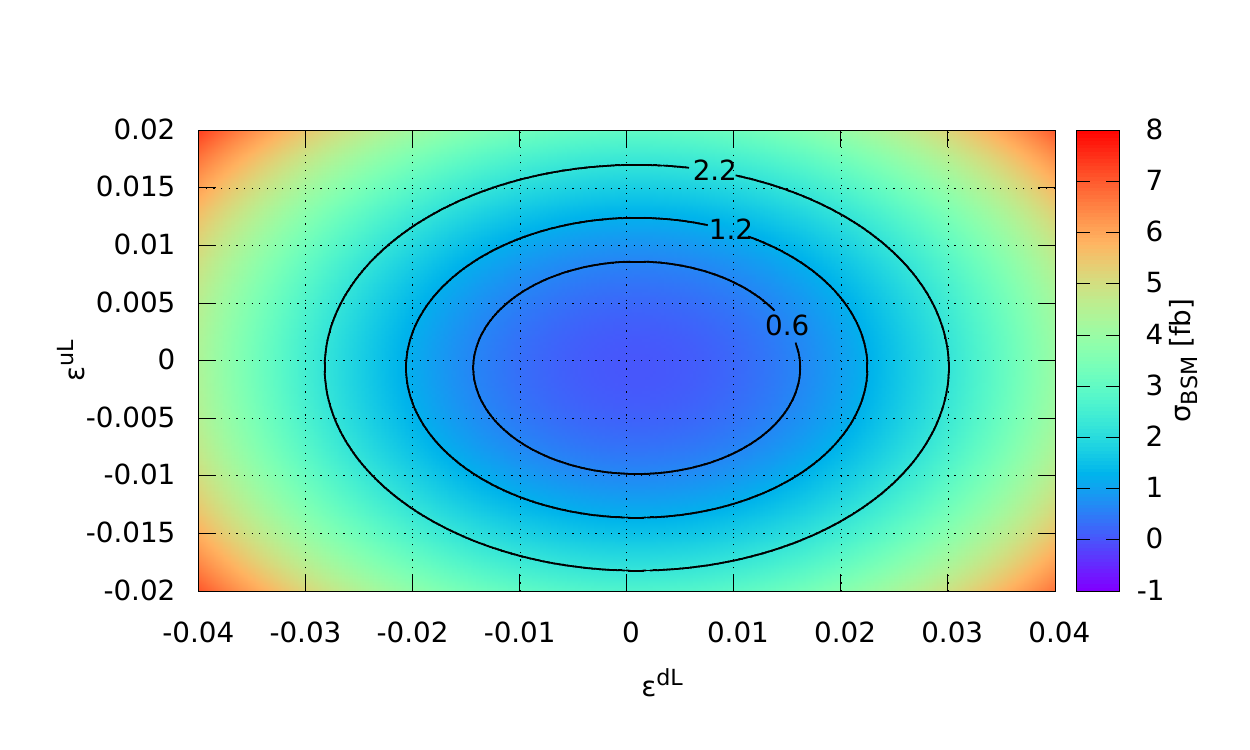}
\vskip -25pt
\includegraphics[angle =0, width=0.495\textwidth, height=0.3\textwidth]{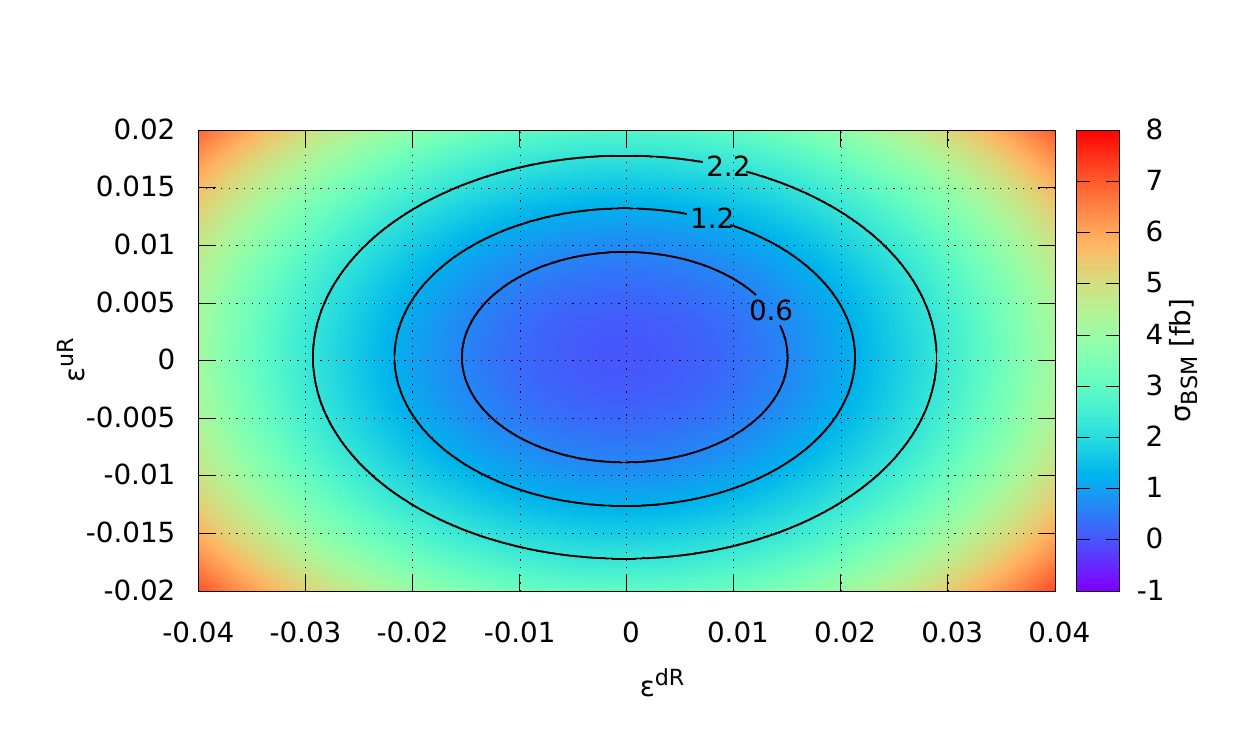}
\end{center}
\vskip -20pt
\caption{NSI contributions (shown by color gradient) to the ATLAS
  search regions for monojet+$p_T\!\!\!\!\!/~$ signature in SR-IM9 as
  a function of $\epsilon^{uL}$ and $\epsilon^{dl}$ (upper panel) and
  $\epsilon^{uR}$ and $\epsilon^{dR}$ (lower panel).}
\label{fig:udl_udr}
\end{figure}

For each of these signal regions, ATLAS collaboration \cite{ATLAS-CONF-2017-060,
  Aaboud:2017phn} have measured the cross sections with 36.1fb$^{-1}$ data 
of 13TeV LHC and provided
 95\% CL upper limits ($\sigma_{\rm obs}^{95}$, also shown in
Table~\ref{monojet_selection}) on the contributions from generic NP
scenarios. In the present context, these could be translated to
ellipsoids in the $\epsilon$-space. Limiting ourselves to a single
pair of operators (we restrict ourselves to $ q = u, d$) at a time,
the NSI contributions (denoted by $\sigma_{\rm BSM}$) corresponding to
the signal region IM9 is illustrated in Fig.\ref{fig:udl_udr}.

The excellent agreement between the numbers of
events observed by the ATLAS detector and that expected within the SM
can be used to impose limits on the parameters $\epsilon_{\alpha
  \beta}^{qX}$ . As pointed out earlier, the final state is
independent of the neutrino flavours, and indeed receives
(incoherently adding) contributions from all possible flavour
combinations. The ensuing constraint, can be parametrized as
\begin{equation}
\sum_{\alpha \beta} \sum_X 
\left[ a_u^{-2} \, \left(\epsilon^{uX}_{\alpha \beta}-\epsilon^{X}_{u}\right)^2
+a_d^{-2} \, \left(\epsilon^{dX}_{\alpha \beta}-\epsilon^{X}_{d}\right)^2 \right]<1,
\label{constraint}
\end{equation}
where the central values  {$\epsilon^{X}_{q}$} are as in
Table~\ref{tab:results}.   {It should be noted that, for off-diagonal 
couplings, $\epsilon^{X}_{q} = $ identically.} 
That $a_u < a_d$ for each case can be
understood as a consequence of the larger densities for the
$u$-quark. Similarly, the fact that $a_{u,d}$ are independent of the
chirality is but reflective of the fact that, in the limit of
vanishing quark masses, terms proportional to $\epsilon^2$ are
independent of chirality.  {The terms linear in
  $\epsilon^{qX}_{\alpha\beta}$} are consequences of interference with
the SM amplitude, {signs being indicative of constructive or
  destructive nature}.  The interferences are large for {the}
left-chiral quarks because of their enhanced coupling with the
$Z$-boson. In fact, for {the $d_L$,} the interference is as
significant as 50\% ($\sigma_{\rm int}^{qX}/\sigma_{\rm
  BSM}^{qX}=2\epsilon_q^X(\epsilon_{\alpha\beta}^{qX}-2\epsilon_q^X)^{-1}$)
for SR-IM1 in the region ($\epsilon_{\alpha\beta}^{dL}\sim 0.02$)
sensitive to LHC run-II data. However, demanding harder cuts on
$\cancel E_T$ reduce the interference contribution significantly.
Note that not much should be read into the nonzero central values
$\epsilon^{X}_{q}$ as these are attributable to small statistical
fluctuations in the data.
\begin{table}[h]
\begin{center}
\begin{tabular}{||c|c|c|c|c|c|c||c|c||c|c||}
\hline \hline
   & & & & & \multicolumn{2} {|c||} {Observed} & \multicolumn{4} {|c||} {Expected}\\
SRs & $\epsilon_u^{L}$ & $\epsilon_u^{R}$ & $\epsilon_d^{L}$ & $\epsilon_d^{R}$ &  \multicolumn{2} {|c||} {36.1 fb$^{-1}$} &  \multicolumn{2} {|c||} {100 fb$^{-1}$} & \multicolumn{2} {|c||} {{300} fb$^{-1}$}\\
\cline{6-11}
 &  &  &  &  & $a_u$ & $a_d$ & $a_u$ & $a_d$& $a_u$ & $a_d$\\
\hline\hline
IM1 & $-23$ & $11$ & $33$ & $-6$ & $564$ & $823$& $311^{69}_{44}$ & $453^{100}_{65}$ & $221^{50}_{47}$ & $323^{72}_{68}$\\ 
IM2 & $-20$ & $9$ & $28$ & $-5$ & $488$ & $716$& $261^{59}_{57}$ & $384^{87}_{84}$ & $186^{42}_{39}$ & $273^{61}_{57}$\\ 
IM3 & $-17$ & $8$ & $24$ & $-5$ & $416$ & $616$& $220^{49}_{49}$ & $325^{72}_{73}$ & $156^{34}_{32}$ & $231^{51}_{48}$\\ 
IM4 & $-15$ & $8$ & $21$ & $-4$ & $334$ & $499$ & $184^{38}_{40}$ & $275^{57}_{59}$ & $131^{29}_{25}$ & $195^{42}_{38}$\\ 
IM5 & $-12$ & $7$ & $16$ & $-3$ & $294$ & $451$& $142^{37}_{32}$ & $219^{57}_{49}$ & $103^{21}_{21}$ & $157^{33}_{32}$\\
IM6 & $-10$ & $5$ & $14$ & $-3$ & $253$ & $397$& $128^{26}_{26}$ & $201^{41}_{41}$ & $92^{18}_{18}$ & $145^{29}_{28}$\\
IM7 & $-8$ & $4$ & $12$ & $-2$ & $207$ & $329$& $128^{26}_{24}$ & $202^{41}_{38}$ & $92^{17}_{17}$ & $146^{26}_{26}$\\
IM8 & $-7$ & $3$ & $10$ & $-2$ & $209$ & $340$& $127^{26}_{23}$ & $206^{41}_{38}$ & $90^{18}_{16}$ & $146^{29}_{26}$\\
IM9 & $-6$ & $3$ & $9$ & $-2$ & $176$ & $292$& $130^{25}_{24}$ & $215^{41}_{39}$ & $92^{21}_{17}$ & $153^{34}_{28}$\\
IM10 & $-5$ & $3$ & $8$ & $-2$ & $187$ & $313$& $133^{23}_{28}$ & $221^{38}_{46}$ & $94^{21}_{12}$ & $157^{35}_{20}$\\
\hline\hline
\end{tabular}
\end{center}

\caption{The  values  of the parameters (each scaled
up by a factor of $10^4$) as in eq.(\ref{constraint}).  {Also shown 
are the expected sizes of the ellipses assuming that the 
agreement of the data with the SM persists with higher luminosities.}}
\label{tab:results}
\end{table}

 {We also show, in Table \ref{tab:results} how $a_{u,d}$ would scale
with luminosity if the present level of agreement between the data and
the SM expectations were to continue. A crucial component in making
this comparison are the systematic uncertainties in the background
estimation.  Listed in Ref.\cite{Aaboud:2017phn}, the dominant
contributions to the uncertainty in the mono-jet background estimation
arise from $(i)$ the uncertainties in the absolute jet and missing
transverse energy scales, $(ii)$ those related to jet quality
requirements, the description of the pileup, b-tagging, lepton
identification and reconstruction efficiency, $(iii)$ those in the
modelling of parton-showers and choice of PDFs, and finally $(iv)$ the
lack of higher-order parton level calculations or the implementation
thereof in the MC event generators.
With increasing amount of data, and hence, a better
understanding of the detector responses, the experimental
uncertainties in the estimation of the SM backgrounds are expected to
be reduced significantly.
As an example, the systematic uncertainty in the background estimation in 
Ref.\cite{Aaboud:2017phn} has reduced nearly by a factor of 2 
when compared to an earlier identical analysis~\cite{Aaboud:2016tnv}
performed with only $3.2\fb^{-1}$ of data. In this even, the indicative 
projections of Table \ref{tab:results} assume that the experimental 
systematic uncertainty would be reduced by a
factor of 2 (4) with accumulated luminosity of $100 (300) \fb^{-1}$.}

It is worthwhile to note that while strengthening the requirement on
$\cancel E_T$ increases the sensitivity (a reflection of the
higher-dimensional nature of the terms), this flattens out at $\cancel
E_T~\gtap~600~\gev$ with the SRs IM6--10 being almost equally
efficient. And in the high-$\cancel E_T$ region, with the semi-axes
$a_{u,d}$ being much larger than $\epsilon^X_q$,
it is the former that essentially determine the shape and size of the
constraint ellipsoids (or, ellipses, when projected to a plane).

\begin{figure}[t]
\begin{center}
\includegraphics[angle =0, width=0.495\textwidth, height=0.25\textwidth]{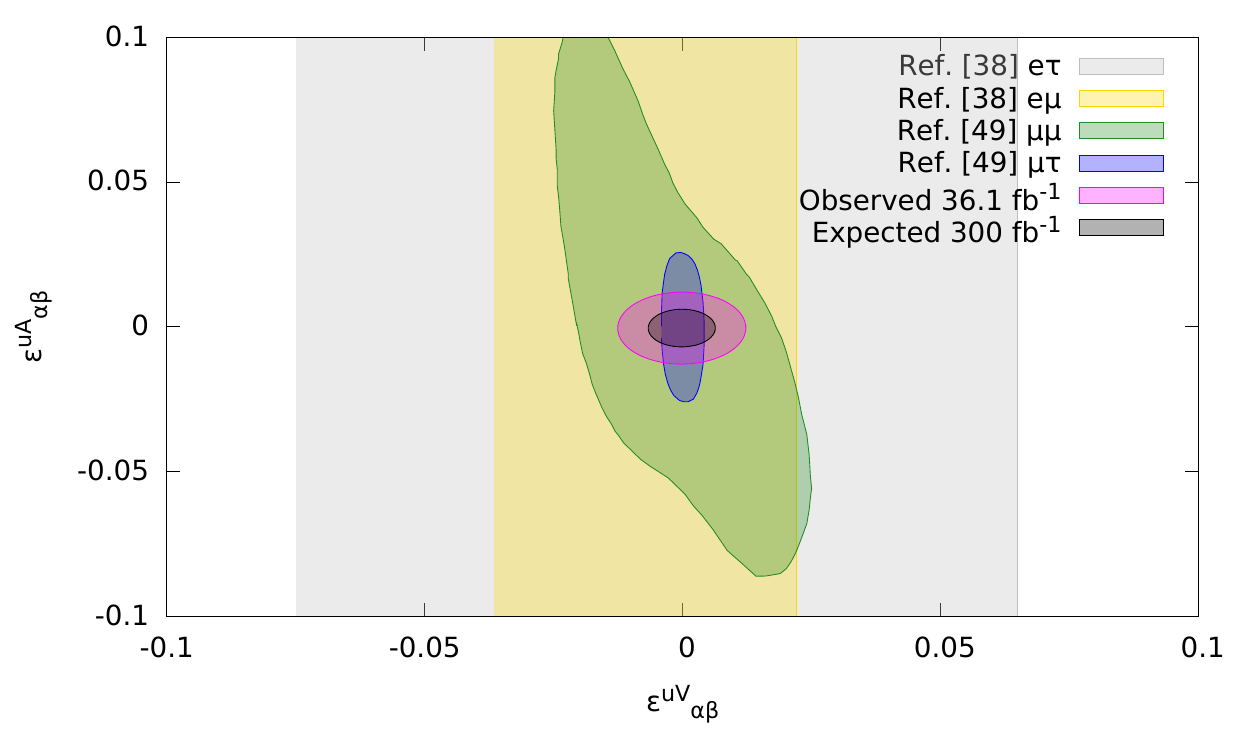}
\includegraphics[angle =0, width=0.495\textwidth, height=0.25\textwidth]{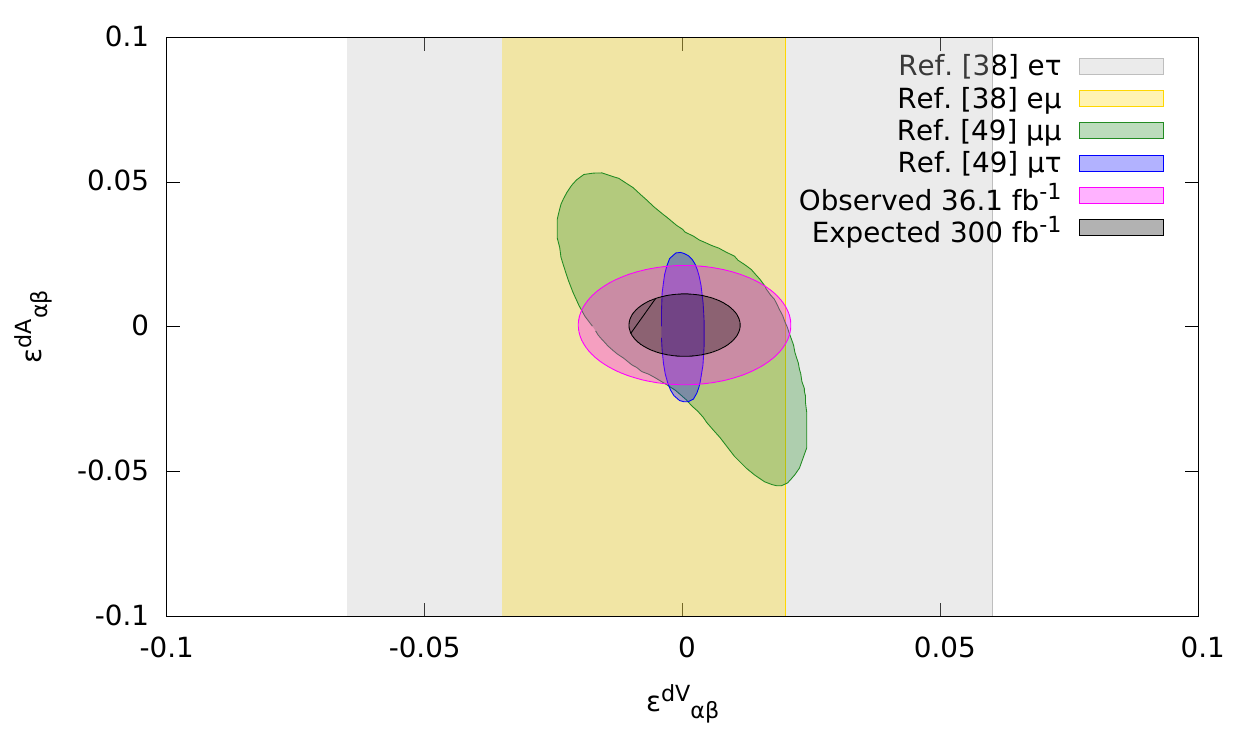}
\end{center}
\caption{Allowed parts of $\epsilon^{qV}_{\alpha
    \beta}$--$\epsilon^{qA}_{\alpha \beta}$ planes, with $q = u (d)$
  in the upper(lower) panels. The bounds from neutrino experiments are
  flavour specific and at 90\% CL, while those from 
    {the ATLAS ($36\fb^{-1}$ at $\sqrt{s} = 13 \tev$)}
 apply to all flavour combinations,  and at 95\% CL.}
\label{fig:bound}
\end{figure}

In Fig.\ref{fig:bound}, we present a comparison of our bounds with
those emanating from other experiments. For neutrino scattering
(whether forward or otherwise) off nonrelativistic nuclei, vector
quark currents contribute more than axial ones, and neutrino
oscillation experiments are only sensitive to $\epsilon^{V}$. Choosing
to work in this basis, 90\% CL bounds on $\epsilon^{fV}_{e\mu}$ and
$\epsilon^{fV}_{e\tau}$ result from a global analysis
\cite{Gonzalez-Garcia:2013usa,Coloma:2017ncl} of data from solar,
atmospheric (Super-Kamiokande \cite{Wendell:2010md}), long-baseline
accelerator experiments (MINOS \cite{Adamson:2013whj,Adamson:2013ue},
T2K \cite{Abe:2013hdq,Abe:2011sj} and reactor experiments (KamLAND \cite{Gando:2010aa},
CHOOZ \cite{Apollonio:1999ae}, Palo Verde \cite{Piepke:2002ju}, Daya
Bay \cite{An:2013uza}, Reno \cite{Ahn:2012nd}). The bounds on
$\epsilon^{fV}_{\mu\mu}$ and $\epsilon^{fV}_{\mu\tau}$ corresponds to
the global analysis \cite{Escrihuela:2011cf} of NuTeV
\cite{Zeller:2001hh} CHARM \cite{Allaby:1987vr}, CDHS
  \cite{Blondel:1989ev} and atmospheric neutrino oscillation
data. Note that oscillation experiments are  sensitive  {only
to off-diagonal $\epsilon$'s and} to
  differences between the diagonal terms (for instance,
  $\epsilon_{ee}-\epsilon_{\mu\mu}$ and
  $\epsilon_{\tau\tau}-\epsilon_{\mu\mu}$). Recently, coherent
  neutrino-nucleus scattering has been observed for the first time by
  the   {{\sc coherent}} experiment \cite{Akimov:2017ade}, 
   {allowing for the derivation of} 
competitive constraints on each of the diagonal parameters
  separately \cite{Coloma:2017ncl}. This is particularly relevant for
  $\epsilon_{ee}^{qV}$ and $\epsilon_{\tau\tau}^{qV}$ for which new
  90\% CL bounds are:
  $-0.045(0.037)<\epsilon_{\tau\tau}^{u(d)V}<0.19(0.16)$ and
  $0.024(0.015)<\epsilon_{ee}^{u(d)V}<0.30(0.27)$.


It is obvious that, for $u$-quark currents, the constraints from the
LHC results are significantly stronger than those from neutrino
experiments, while those for the $d$-quark currents are more than
competitive.  At this stage, let us reexamine the NSI operators in
totality.  Since these, presumably, owe their origin to physics beyond
the SM, the operators in eq.(\ref{4fermi}) ought to be written in
terms of $SU(2)_L \otimes U(1)_Y$ invariant terms. For a pair of
lepton doublets $L_{\alpha, \beta}$, the triplet combination would
introduce terms of the form $(\bar \nu_\alpha \gamma_\mu P_L
\ell_\beta) \, (\bar d \gamma^\mu P_L u)$ leading to extra
contributions to well-measured meson decays (or, the decay of a $\tau$
to a meson). In the context of the LHC, on the other hand, this would
lead to lepton nonuniversality in $ p p \to \ell + \nu $ (exclusive or
inclusive). Both sets of observables would lead to constraints
much stronger than those discussed above. On the other hand, were 
we to consider a singlet structure, namely,
\[
  \bar L_\alpha \gamma_\mu L_\beta 
     = \bar \nu_\alpha \gamma_\mu \nu_\beta + 
     \bar \ell_\alpha \gamma_\mu \ell_\beta \ ,
\] 
clearly $\alpha \neq \beta$ would lead to $ p p \to \ell_\alpha +
\bar\ell_\beta $.  Vetoing events with substantial missing energy
(thereby suppressing the $WW$ background) would lead to spectacular
signals for $\epsilon_{\alpha\beta}$ being considered here. Even stronger
bounds would emanate from lepton flavour changing decays of neutral
mesons.

For $\alpha=\beta$, the {charged lepton} bounds are, understandably,
weaker. However, even in this case, high-mass dilepton ($e^\pm$ or
$\mu^\pm$) production constrains 4-fermi operators to a contact
interaction scale of $\gtap~25~\tev$~\cite{Aaboud:2017buh}. Translated
to our language, this would imply $\epsilon^{u,d}_{ee},
\epsilon^{u,d}_{\mu\mu}~\ltap~5\times 10^{-5}$.  In principle, even
stronger bounds can be obtained by considering
asymmetries~\cite{Choudhury:2002av}. It would seem, thus, that
$\epsilon^{u,d}_{\tau\tau}$ are the only Wilson coefficients {(for
  first-generation quarks)} that {would have} remained significantly
unconstrained by either low-energy observables or LHC observables such
as dilepton production.  However, our analysis improves the situation
dramatically, and far supersedes the   {{\sc
    coherent}} bounds~\cite{Coloma:2017ncl}. {Note, furthermore,
  that we have not included the CMS data yet, which too does not show
  any excess in the monojet signal. However, with the CMS typically
  imposing softer requirements on both the leading jet and $\cancel
  E_T$, the reported exclusion~\cite{Sirunyan:2017jix} of $\sigma_{\rm BSM}$ is
  weaker than that in Ref.~\cite{Aaboud:2017buh}.  Once CMS reanalyses
  their data, the ensuing constraints can be combined with those
  reported here to  yield significantly stronger bounds.}

The narrative would change were we to consider suppressing operators
involving the charged lepton by means of postulating multiple $SU(2)_L
\otimes U(1)_Y$ invariant operators with carefully tuned
WCs~\cite{Biggio:2009nt}. A different approach would be to postulate
dimension-8 operators such as $\left[(\phi^* \bar L_\alpha) \gamma_\mu
  (\phi L)\right] \, (\bar q \gamma^\mu q)$ where $\phi$ is the SM
Higgs doublet. In either case, charged-lepton 4-fermion operators do
not exist, and the low-energy constraints are rendered very weak.
Similarly, the simplest collider constraints are not operative either,
and while more exotic signatures are suddenly possible, the
corresponding cross sections are too small to be of any interest with
the currently accumulated luminosity. However, as we have
 {conclusively established} in this article,
even these scenarios  {(and any variants thereof)} are already
severely constrained by a simple final state such as a monojet with
missing energy. And with the luminosity that the LHC is slated to
deliver, continuing negative results would only strengthen the
constraint to well beyond what even a next-generation neutrino
experiment will be able to probe \cite{Choubey:2015xha,Choubey:2014iia,Chatterjee:2014gxa,Fukasawa:2016lew,Ohlsson:2012kf}. This would indicate that the only
role such facilities may play in this regard would be the confirmatory
one. 

DC acknowledges partial support from the European Union’s Horizon 2020
research and innovation programme under Marie Sk{\l}odowska-Curie
grant No 674896, and the R\&D grant of the University of Delhi. KG is
supported by  {the} DST (India) under INSPIRE Faculty Award. SN
acknowledges the hospitality of  {the} Institute of Physics,
Bhubaneswar during  {the} initial phase of the work.

\bibliographystyle{apsrev}
\bibliography{reference}
\end{document}